\RequirePackage[T1]{fontenc}
\documentclass[conference,a4paper]{IEEEtran}
\IEEEoverridecommandlockouts
\usepackage{cite}
\usepackage{amsmath,amssymb,amsfonts}
\usepackage{algorithmic}
\usepackage{graphicx}
\usepackage{textcomp}
\usepackage{xcolor}

\usepackage{multirow}
\usepackage[a4paper, total={184mm,239mm}]{geometry}
\usepackage{fancyhdr}
\fancypagestyle{icecsheader}{%
    \fancyhf{}
    \fancyhead[C]{%
        \normalcolor\fontsize{8}{9}\selectfont
        2026 IEEE 33rd International Conference on Electronics, Circuits and Systems (ICECS)}

}
\fancypagestyle{icecsfirstpage}{%
    \fancyhf{}
    \fancyhead[C]{%
        \normalcolor\fontsize{8}{9}\selectfont
        2026 IEEE 33rd International Conference on Electronics, Circuits and Systems (ICECS)}
    \fancyfoot[L]{%
        \normalcolor\fontsize{8}{9}\selectfont
        979-8-3195-1905-4/26/\$31.00~\textcopyright{}~2026 IEEE}

}
\def\BibTeX{{\rm B\kern-.05em{\sc i\kern-.025em b}\kern-.08em
    T\kern-.1667em\lower.7ex\hbox{E}\kern-.125emX}}
\usepackage{todonotes}
\renewcommand{\baselinestretch}{0.955}

\usepackage{tikz}

\DeclareRobustCommand{\figtag}[1]{%
\raisebox{0.18ex}{%
\tikz[baseline=(char.base)]{%
\node[
circle,
fill=black,
text=white,
inner sep=0pt,
minimum size=1.25em,
font=\scriptsize\bfseries
] (char) {#1};%
}%
}%
}

\begin{document}

\title{FSNIC: A Low-Latency Flow-Based Intrusion Detection Architecture for FPGA SmartNICs}
\author{
Nise O'Cuill,
Changhong Li,
Georgios Floros,
Shreejith Shanker \\ 
Reconfigurable Computing Systems Lab, Electronic \& Electrical Engineering\\
Trinity College Dublin, Ireland\\
Email: \{ocuilln, lic9, florosg, shreejith.shanker\}@tcd.ie
}
\maketitle
\IEEEpubid{%
    \makebox[\columnwidth][l]{%
        \fontsize{8}{9}\selectfont
        979-8-3195-1905-4/26/\$31.00~\textcopyright{}~2026 IEEE}%
    \hspace{\columnsep}%
    \makebox[\columnwidth]{}}
\thispagestyle{icecsfirstpage}

\begin{abstract}

Modern data centres require high-performance networking alongside effective real-time security.
Traditional Intrusion Detection Systems (IDS) commonly rely on general-purpose processors and often struggle to inspect high-speed traffic at line rate without introducing latency or performance bottlenecks.
Smart Network Interface Cards (NICs) provide an alternative by enabling computation directly within the network data plane.
This work presents a machine learning-based IDS implemented within an FPGA-based SmartNIC pipeline.
The system integrates P4-based packet parsing with a LogicNets IDS model implemented in RTL, enabling deterministic, low-latency inference.
Compared with traditional stateless packet-level classifiers, the proposed stateful flow-based IDS introduces minimal state by aggregating features across packets, capturing behavioural patterns not observable at the packet level.
Experimental results on the UNSW-NB15 dataset show that the flow-based IDS improves detection accuracy from 86.92\% to 97.68\% compared with stateless packet-level classification.
We also evaluate the proposed IDS on CICIDS2017 and compare its real-time hardware performance with prior FPGA-based IDS designs. Through hardware-software co-design, the proposed IDS achieves 6~ns inference latency using only 846 LUTs, with no BRAM or DSP usage, demonstrating a low latency and resource efficient implementation.

\end{abstract}

\section{Introduction}\label{introduction}

Increasing bandwidth demands on modern data centres, combined with growing CPU overhead, have led to a significant interest in moving computation closer to the network interface through compute-in-network architectures~\cite{doring2021smartnics, kfoury2024smartnic}.
SmartNICs provide programmable packet-processing capabilities directly within the network data plane, enabling functions such as monitoring, filtering and intrusion detection to be executed without interaction with the host.
This reduces the host overhead while enabling deterministic, low-latency packet processing.
However, existing machine learning-based intrusion detection system (IDS) approaches are poorly aligned with programmable data-plane constraints.

Models that achieve high detection accuracy typically rely on computationally expensive temporal architectures, while lightweight in-network approaches often lack sufficient behavioural context for reliable intrusion detection.
Existing software-based IDS approaches frequently rely on deep learning architectures such as LSTMs to capture temporal traffic behaviour and achieve high detection accuracy \cite{sayegh2024lstmids}.
While effective, these approaches introduce sequential dependencies and variable processing latency that are difficult to accommodate within FPGA-based SmartNIC environments.

In contrast, lightweight in-network IDS approaches implemented directly within programmable data planes satisfy strict latency requirements but often operate using stateless packet-level features \cite{reddy2023p4, sada2025innetworkml}.
Although this enables line-rate execution, the lack of temporal and behavioural context limits the ability to distinguish between benign and malicious traffic exhibiting similar packet-level characteristics.
Many network attacks exhibit behavioural characteristics that emerge only across sequences of packets, making reliable classification difficult using isolated packet-level features alone.
This trade-off highlights a fundamental challenge in programmable data plane intrusion detection: how to incorporate sufficient temporal context without violating the deterministic execution and resource constraints of SmartNIC hardware.
As a result, current programmable data plane IDS systems typically sacrifice either detection capability or deployability, with few approaches achieving both high accuracy and deterministic hardware-efficient execution.

Machine-learning-based IDS systems have often utilised custom compilation using frameworks such as FINN~\cite{umuroglu2017finn} to achieve high throughput and line-rate detection~\cite{le2022feature}.
However, these designs typically rely on arithmetic operations and multi-stage pipelines, increasing hardware complexity and latency.
Logic-based neural networks such as LogicNets~\cite{umuroglu2020logicnets} provide an alternative approach by implementing inference directly as LUTs, enabling deterministic low-latency execution without DSP or memory overhead.
LUT-based IDS accelerators have achieved promising real-time performance, but existing designs typically remain standalone packet-level classifiers and do not exploit flow-level behavioural context within the SmartNIC data path~\cite{farooq2025high}.


To bridge these gaps, this work presents a hardware-efficient, flow-based IDS architecture for programmable SmartNIC deployment using lightweight temporal aggregation and LogicNets inference.
The proposed architecture aggregates lightweight flow statistics directly within the data plane, enabling behavioural context to be captured while maintaining deterministic low-latency execution.

The main contributions of this work are as follows:


\begin{itemize}
    \item A hardware-efficient flow-based IDS architecture for programmable SmartNIC deployment using lightweight temporal aggregation and LogicNets inference.

    \item A packet reconstruction pipeline for deriving temporally ordered flow-window representations from UNSW-NB15 and CICIDS2017 PCAP traffic traces, enabling programmable data-plane evaluation.

    \item An evaluation on UNSW-NB15 and CICIDS2017 demonstrates that lightweight temporal feature aggregation improves detection performance on UNSW-NB15 and maintains strong classification performance across both datasets with deterministic low latency execution and low FPGA resource utilisation.
\end{itemize}

\section{Methodology and Design} \label{methodology}
\subsection{Overall SmartNIC IDS Framework}
\begin{figure}[t]
    \centering
    \includegraphics[width=\columnwidth]{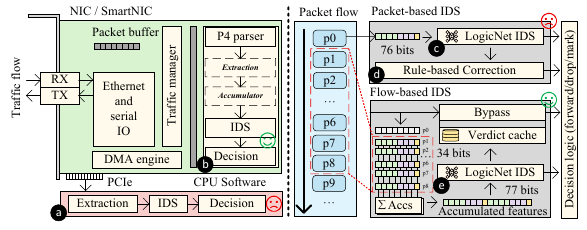}
    \caption{Overall SmartNIC IDS architecture with host software, packet-based, and flow-based detection paths.}
    \label{fig:overall_framework}
    \vspace{-5mm}
\end{figure}

Fig.~\ref{fig:overall_framework} presents the overall framework of the proposed SmartNIC IDS compared with conventional approaches.
For a conventional software IDS workflow, which is demonstrated as \figtag{a},
packets are transferred from the NIC DMA engine to the CPU through the PCIe interface, where feature extraction and IDS classification are performed in software.
Although this workflow is flexible for offline analysis and software prototyping, it requires host intervention and data movement across the PCIe interface, which can become a bottleneck for line-rate intrusion detection.

The SmartNIC pipeline, marked by \figtag{b}, moves packet parsing, feature extraction, and IDS decision logic into the data path.
Incoming traffic is temporarily stored in the packet buffer, and the traffic manager schedules packets for P4-based header parsing and feature extraction.
The IDS decision is delivered to the decision logic, which determines whether the traffic should be forwarded, dropped, marked, or passed to the host.
On the SmartNIC, IDS classification and flow control updates are performed directly in the data path, avoiding host software intervention and PCIe data movement.
For the conventional packet-level IDS demonstrated in~\figtag{c}, each packet is independently converted into a 76-bit feature vector by the P4 parser and classified by the hardware IDS.
Moreover, in some approaches, a rule-based correction will also be applied to refine the classifier output using selected protocol-level conditions as shown in~\figtag{d}.
This branch provides immediate per-packet inference and therefore introduces minimal decision delay.
However, the classifier only observes instantaneous packet-level information, so it cannot directly capture behavioural patterns that emerge across multiple packets in the same flow.

In the proposed flow-based IDS, as shown in~\figtag{e}, eight consecutive packets are grouped into a compact flow window. The eight-packet window was selected as a simple design choice for the proof-of-concept implementation, providing short-term temporal context while maintaining a fixed and lightweight aggregation structure. The architecture is not restricted to this window size and can be extended to aggregate a larger number of packets. 
For each packet, the P4 parser produces 34-bit metadata for the flow accumulator. Packet count, length statistics, direction and TCP flag distributions form a 77-bit feature vector for the customised IDS. The resulting verdict is cached and reused for subsequent packets of the same flow, avoiding redundant inference. The following subsections detail feature construction and SmartNIC integration.

\subsection{P4-Compatible Stateful Flow Feature Construction}





\begin{figure}[t]
\centering
\includegraphics[width=0.98\columnwidth]{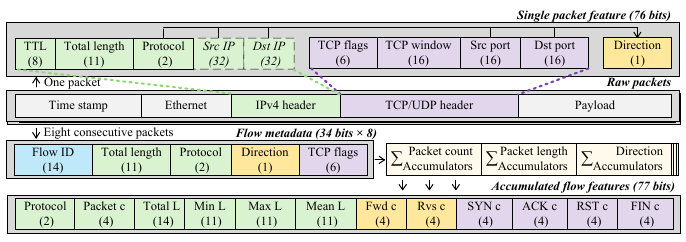}
\caption{P4-compatible packet parsing and stateful flow-window feature construction for SmartNIC-based intrusion detection.}
\label{fig:p4_flow_feature_construction}
\end{figure}

Programmable SmartNIC intrusion detection starts with raw Ethernet frames arriving at the data plane.
As illustrated in Fig.~\ref{fig:p4_flow_feature_construction}, each packet is parsed into Ethernet, IPv4 and TCP/UDP headers, while the payload is ignored to preserve line rate processing.
The P4 parser extracts only deterministic header fields that can be obtained without payload inspection or software side flow reconstruction.

The conventional packet-level baseline uses a 76-bit single packet feature vector.
It is extracted independently from each incoming packet and consists of TTL, total length, protocol, TCP flags, TCP window size, source port, destination port and packet direction.
In stateless packet classification, each packet is treated as an isolated observation, so the classifier can capture instantaneous header information but cannot observe how the flow evolves across multiple packets.

In our proposed SmartNIC, the parsed packet is first associated with a flow using a five-tuple key formed by source IP address, destination IP address, source port, destination port and transport protocol.
The five-tuple is used for flow identification and state indexing.
After the flow entry is identified, the packet is compressed into a 34-bit flow metadata vector, consisting of a 14-bit flow ID, an 11-bit packet length field, a 2-bit protocol field, a 1-bit direction flag and a 6-bit TCP flag field.

The 34-bit metadata is the per-packet update input to the RTL flow accumulator in order to retain the minimum information needed for temporal aggregation while removing fields that are unnecessary for repeated arithmetic updates.
For each packet belonging to the same flow, the accumulator updates packet count, total length, minimum length, maximum length, mean length, forward and reverse packet counts, and TCP flag counters. This process is repeated over eight consecutive packets from the same flow.

The final input to the proposed classifier is a 77-bit accumulated flow feature vector.
It includes protocol, packet count, total packet length, minimum length, maximum length, mean length, forward packet count, reverse packet count, and SYN, ACK, RST and FIN counters.
Compared with the 76-bit packet feature, the 77-bit feature represents short-term flow behaviour over an eight packet window.
The features enable stateful flow-aware inference using compact temporal statistics, while retaining compatibility with P4-controlled parsing and FPGA-based SmartNIC execution.

\subsection{SmartNIC Coupled Flow-Level IDS Architecture}
Within the packet engine, incoming packets are parsed in the pipeline to extract header-level features and generate flow identifiers.
Packets belonging to the same flow are accumulated in a lightweight flow state structure, implemented as FIFO based shift register logic parallel to the data path.
The aggregated feature vectors corresponding to the latest eight packets from the flow are forwarded to the intrusion detection engine.
Following classification, the resulting verdict is cached and applied to subsequent packets belonging to the same flow.
Unlike stateless packet-level IDS, this approach captures flow dynamics through the lightweight temporal aggregation within the dataplane, which allows behavioural properties such as directional traffic patterns, packet-size variation and TCP flag distributions to be captured without significant hardware overhead. 
P4's \texttt{extern} interface is used to couple the IDS engine in the design (mapped as AXI4 interfaces in the final hardware). 
After functional validation, the P4 pipeline is compiled into the hardware model using AMD's Vitis Networking P4 (VNP4) framework, targeting an Alveo U280 FPGA (\texttt{xcu280-fsvh2892-2L-e}).

For our IDS engine, we started with the NID-S architecture from LogicNets~\cite{umuroglu2020logicnets} as the baseline design, and subsequently fine-tuned the model through design space exploration. 
The 77--593--100--1 classifier had two hidden layers with 593 and 100 neurons, and used quantised ReLU activations in the hidden layers and a quantised HardTanh activation at the output. Quantisation-aware training in Brevitas used 1-bit inputs and 2-bit hidden and output activations, with a fixed input scale and learned hidden and output scales.
For each neuron node, a fixed input degree $\gamma=7$ is applied to implement the sparse connectivity and reduce the LUT utilisation.
After quantisation and sparsification, the model is translated into layer-wise RTL logic blocks. Pipeline registers are inserted between layers, avoiding FIFO-based inter-layer buffering commonly used in FINN-style IDS accelerators~\cite{le2021towards, le2022feature}, reducing inference latency.

The classification result from the model is passed back into the pipeline through the AXI4 interface for downstream processing (drop/tag packets in the flow). 
As mentioned before, the result is also cached and applied to subsequent packets belonging to the same flow to overlap inference with packet reception.
Once integrated, the design is compiled using AMD's Vitis/Vivado tools to generate the final integrated design for the target platform.

\section{Experimental Results}
\label{results}
\subsection{Experimental Setup}
The P4 data-plane implementation, model training and LogicNets-based RTL accelerator generation were performed on a workstation equipped with an Intel Core i7-10700K CPU and an NVIDIA GeForce RTX 3080 Ti GPU.
The software environment uses Python 3.12, Vitis 2024.2 and Vitis Networking P4.
Resource utilisation and latency results are obtained from out-of-context synthesis and RTL simulation reports using the Vivado/Vitis 2024.2 toolchain, targeting an Alveo U280 FPGA (\texttt{xcu280-fsvh2892-2L-e}). The reported 500~MHz frequency refers specifically to the out-of-context IDS engine implementation; the operating frequency of a complete SmartNIC implementation is dependent on the target platform and surrounding data-path infrastructure.
The evaluation uses reconstructed packet traces derived from the UNSW-NB15 and CICIDS2017 datasets and RTL simulation rather than live traffic on a deployed SmartNIC. This enables controlled functional evaluation of the proposed architecture; full board-level validation under live traffic is proposed for future work.

\subsection{Flow-based IDS Performance}
\begin{figure}[t]
\centering
\includegraphics[width=0.95\columnwidth]{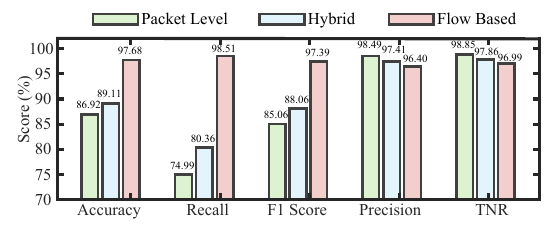}
\caption{Comparison of intrusion detection architectures on UNSW-NB15.}
\label{fig:architecture_comparison}\vspace{-2mm}
\end{figure}

Fig.~\ref{fig:architecture_comparison} compares comprehensive IDS classification performance of stateless packet-level, hybrid and flow-based intrusion detection architectures evaluated on the UNSW-NB15 dataset.
Although the packet-level architecture achieves slightly lower false-positive rates, it suffers from significantly reduced detection capability due to the absence of temporal information.
The hybrid architecture provides moderate improvement over stateless packet-level classification, but still underperforms the proposed flow-based approach.
The proposed flow-based architecture improves detection capability by incorporating lightweight temporal aggregation across eight sequential packets, and its accuracy and F1 score significantly outperform the other two approaches, indicating that temporal behavioural context is critical for distinguishing malicious traffic patterns that are not observable from isolated packet-level features alone.

\begin{table}[t!]
\centering
\caption{Flow-Based IDS Evaluation Across Datasets}
\label{tab:flow_dataset_detailed}
\begin{tabular}{lcc}
\hline
Metric & UNSW-NB15 & CICIDS2017 \\
\hline
Accuracy & 97.68\% & 96.40\% \\
Recall & 98.51\% & 92.86\% \\
TNR & 96.99\% & 99.94\% \\
Precision & 96.40\% & 99.93\% \\
F1 Score & 97.45\% & 96.27\% \\
False Positives & 94 & 2 \\
False Negatives & 38 & 223 \\
\hline
\end{tabular}
\end{table}

\begin{table}[t!]
\centering
\caption{Hierarchical Resource Utilisation of the Proposed IDS}
\label{tab:hierarchical_resources}
\begin{tabular}{lcccc}
\hline
Component & LUTs & FFs & BRAM & DSP \\
\hline
Flow accumulator & 62 & 69 & 0 & 0 \\
LogicNet classifier & 784 & 149 & 0 & 0 \\
Top-level/control logic & 0 & 34 & 0 & 0 \\
\hline
Total & 846 & 252 & 0 & 0 \\
\hline
\end{tabular}
\end{table}

\begin{table*}[ht]
\centering
\caption{Comparison with Prior FPGA-Based Intrusion Detection Designs Benchmarked with UNSW-NB15 and CICIDS2017}
\label{tab:ids_comparison}
\scalebox{0.95}{
\setlength{\tabcolsep}{4pt}
\begin{tabular}{lccccccc}
\hline
Work & On-chip Feature Extraction & UNSW-NB15 Acc. (\%) & CICIDS2017 Acc. (\%) &
Latency (ns) & Throughput (PPS) & LUT & $F$ (MHz) \\
\hline
\cite{murovivc2021genetically} & No  & 92.04 & --    & 19        & 500 M & 26,879 & -- \\
\cite{umuroglu2020logicnets}   & No  & 91.30 & --    & 10.5      & --                   & 15,949 & 471 \\
\cite{vrevca2021detecting}     & No  & 82.10 & --    & 91        & 11 M & 26,556 & 142.85 \\
\cite{le2021towards}           & Yes & --    & 99.40 & --        & 176 K & 26,615 & 100 \\
\cite{le2022feature}           & Yes & 98.65 & 97.39 & 1,780,000 & 2.43 K                & 22,893 & 100 \\
\cite{le2022feature}           & Yes & 99.80 & 99.84 & 85,260    & 86.67 K               & 59,679 & 100 \\
\cite{le2022feature}           & Yes & 99.77 & 99.84 & 22,270    & 565 K              & 47,297 & 100 \\
\cite{le2022feature}           & Yes & 99.51 & 99.82 & 820,000   & 165 K             & 45,119 & 100 \\
\cite{le2022feature}           & Yes & 99.77 & 99.84 & 6,680     & 1.9 M            & 47,297 & 336.13 \\
Prop.                          & Yes  & 97.68 & 96.40 & 6         & 500 M                  & 846    & $500^{*}$ \\
\hline 
\end{tabular}} 

\vspace{-3mm}
\end{table*}

We evaluate the P4-based flow-feature pipeline on both datasets. Table~\ref{tab:flow_dataset_detailed} reports strong flow-window classification performance on UNSW-NB15 and CICIDS2017, supporting its applicability beyond a single benchmark.
The CICIDS2017 evaluation achieves a higher precision and TNR than UNSW-NB15, indicating excellent benign traffic discrimination. However, false negatives are particularly important for intrusion detection, as they correspond to malicious flows that remain undetected. The proposed IDS produces 38 false negatives on UNSW-NB15 and 223 on CICIDS2017, corresponding to false-negative rates of 1.49\% and 7.14\%, respectively. The higher false-negative rate on CICIDS2017 indicates that some attack behaviours are less consistently captured by the selected flow statistics, highlighting attack recall as an important target for further optimisation.
Table~\ref{tab:hierarchical_resources} reports the proposed IDS's resource utilisation. Combining sparse classifier connectivity with lightweight RTL accumulation and control keeps the additional hardware overhead small, enabling temporal aggregation without computationally expensive recurrent models.

\subsection{Comparison with Prior Work}
Table~\ref{tab:ids_comparison} compares the accuracy and hardware performance of FPGA-based IDS implementations.
Accelerators in \cite{murovivc2021genetically, umuroglu2020logicnets} and \cite{vrevca2021detecting}, as earlier studies, used BNNs to achieve relatively low resource usage and nanosecond-level inference latency, making them particularly suitable for edge real-time tasks like IDS.
The aggressive binarisation used in BNN-based designs can limit model capacity, which often results in lower accuracy on more complex traffic datasets.
Furthermore, the validation of these BNN models on UNSW-NB15 usually uses pre-binarised datasets, which is often difficult to implement in real-time online feature extraction.
Therefore, most of these methods also adopt offline feature extraction and evaluation modes.
Previous work~\cite{le2021towards} introduced a FINN-based CNN accelerator with on-chip flow-bucket feature extraction. Their subsequent work~\cite{le2022feature} extended this direction by incorporating raw-packet feature extraction and evaluating multiple feature-dimensionality configurations.
While the flow analysis of the original packets and the high-precision FINN model greatly improve the accuracy of the model, latency and resource utilisation also increase significantly.
Our model achieves minimal latency and LUT usage while improving accuracy and generalisation by using raw packet flow features, and benefits from the LogicNets model, lightweight RTL components, and tightly coupled SmartNIC integration.
These results demonstrate a substantially smaller hardware footprint and lower inference latency than the FINN-based configurations in ~\cite{le2021towards, le2022feature}, although direct comparison is affected by differences in model architecture, operating frequency, feature extraction and implementation methodology.
\section{Conclusion}
\label{conclusion}


This paper presents a flow-based intrusion detection architecture tightly coupled with a SmartNIC data path.
The lightweight temporal aggregation enables reliable behavioural intrusion detection with the raw packets while maintaining deterministic low-latency inference and extremely low FPGA resource utilisation, demonstrating its suitability for resource-constrained SmartNIC deployment and cross-dataset evaluation.
Future work will include board-level deployment and evaluation under live network traffic, including high flow-churn conditions, as well as extending the framework towards an Intrusion Prevention System tightly coupled with the SmartNIC data path.


\bibliographystyle{IEEEtran}

\bibliography{references}

\end{document}